\documentclass[prc,twocolumn,showpacs,floatfix,nofootinbib,preprintnumbers,superscriptaddress,amsmath,amssymb]{revtex4}

\usepackage{CJKutf8}
\usepackage{graphicx}
\usepackage{epsfig}
\usepackage{amsmath,amssymb}
\usepackage{bm}
\usepackage{color}
\usepackage{float}
\usepackage{dcolumn}

\begin{document}
\begin{CJK*}{UTF8}{gbsn}

\title{Rotational properties of nuclei around $^{254}$No investigated
 using a spectroscopic-quality Skyrme energy density functional}

\author{Yue Shi (石跃)}
\affiliation{Department of Physics and Astronomy, University of Tennessee, Knoxville, Tennessee 37996, USA}
\affiliation{Joint Institute for Heavy Ion Research, Oak Ridge National Laboratory, Oak Ridge, Tennessee 37831, USA}
\affiliation{Department of Physics, PO Box 35 (YFL), FI-40014 University of Jyv{\"a}skyl{\"a}, Finland}
\affiliation{State Key Laboratory of Nuclear Physics and Technology, School of Physics, Peking University, Beijing 100871, China}

\author{J. Dobaczewski}
\affiliation{Department of Physics, PO Box 35 (YFL), FI-40014 University of Jyv{\"a}skyl{\"a}, Finland}
\affiliation{Institute of Theoretical Physics, Faculty of Physics, University of Warsaw, ul. Ho{\.z}a 69, PL-00681 Warsaw, Poland}

\author{P.T. Greenlees}
\affiliation{Department of Physics, PO Box 35 (YFL), FI-40014 University of Jyv{\"a}skyl{\"a}, Finland}

\begin{abstract}
\begin{description}
\item[Background]
Nuclei in the $Z\approx100$ mass region represent the heaviest
systems where detailed spectroscopic information is experimentally
available. Although microscopic-macroscopic and self-consistent models
have achieved great success in describing the data in this mass
region, a fully satisfying precise theoretical description is still missing.

\item[Purpose]
By using fine-tuned parametrizations of the energy density
functionals, the present work aims at an improved description of the
single-particle properties and rotational bands in the nobelium
region. Such locally optimized parameterizations may have better
properties when extrapolating towards the superheavy region.

\item[Methods]
Skyrme-Hartree-Fock-Bogolyubov and Lipkin-Nogami
methods were used to calculate the quasiparticle energies and rotational
bands of nuclei in the nobelium region. Starting from the most recent Skyrme
parametrization, UNEDF1, the spin-orbit coupling
constants and pairing strengths have been tuned, so as to achieve a better agreement
with the excitation spectra and odd-even mass differences in
$^{251}$Cf and $^{249}$Bk.

\item[Results]
The quasiparticle properties of $^{251}$Cf and $^{249}$Bk were very
well reproduced. At the same time, crucial deformed neutron and
proton shell gaps open up at $N=152$ and $Z=100$, respectively.
Rotational bands in Fm, No, and Rf isotopes, where experimental data
are available, were also fairly well described. To help future
improvements towards a more precise description,  small
deficiencies of the approach were carefully identified.

\item[Conclusions]
In the $Z\approx100$ mass region, larger spin-orbit strengths
 than those from global adjustments lead to improved agreement
with data. Puzzling effects of particle-number restoration on the calculated moment of inertia, at odds with
the experimental behaviour, require further scrutiny.

\end{description}
\end{abstract}


\maketitle
\end{CJK*}

\section{Introduction}
\label{sec1}

The stability of Superheavy elements (SHE) with atomic numbers $Z \ge 104$
is entirely due to quantum shell effects. Without such shell effects, strong Coulomb repulsion
between the protons would lead to immediate spontaneous fission.
Therefore, one of the most important subjects
in the study of SHE is nuclear shell structure. However, due to
low population cross sections, experimental information on SHE
is generally limited to half-lives and $\alpha$-decay energies, which can only
be indirectly related to the shell structure. Only very
recently were electromagnetic transitions observed for the first time in the decay
chain of $^{288}115$ ~\cite{(Rud13)}. After decades
of experimental studies, the heaviest SHE so far produced has proton number
$Z=118$~\cite{(Hof00),(Oga07),(Hof11)}.

Nuclei in the nobelium ($Z=102$) region are unique in the sense that they are
the heaviest systems where detailed spectroscopic information is
available~\cite{(Her08)}. With numerous rotational bands
observed~\cite{(Gre12)}, the ground states of these nuclei are known to be well
deformed, with prolate $\beta_2\approx0.3$ shapes.  Although the SHE in the
elusive superheavy island of stability are thought to be spherical, it can
be expected that a proper description of the single-particle (s.p.) structures in
the nobelium region would result in a better extrapolation of the shell
structure towards the superheavy region.

Theoretical calculations in the nobelium region have been limited to various
mean-field models, which can be divided into two large categories: those based
on the microscopic-macroscopic
model~\cite{(Nil68),(Sob01),(Sob07),(Sob11),(Liu12),(Zha13)}, and those using
self-consistent approaches~\cite{(Ben03)}. The latter used
non-relativistic zero-range Skyrme~\cite{(Cwi96),(Cwi99),(Dug01w),(Ben03d)} or
finite-range Gogny~\cite{(Egi00),(Del06),(War12)} forces as well as
relativistic Lagrangians~\cite{(Afa03),(Vre05b),(Lit12),(Pra12),(Afa13)}. All these
calculations reproduce reasonably well the gross features of experimental
rotational spectra, however, differences can be found on detailed inspection.

Experimental quasiparticle energies of odd-$A$ systems in
the nobelium region show better agreement with the microscopic-macroscopic
calculations~\cite{(Sob11)}, which have deformed shell gaps open at
$N=152$ and $Z=100$. None of the self-consistent approaches, whether
in the non-relativistic or relativistic variant, predict such
deformed shell gaps for both protons and neutrons; a few of them
predict the $N=152$ or $Z=100$ shell openings, but not
both~\cite{(Ben13)}.

The fine details of the shell structure depend on small changes of the model
parameters. Up to now, the parameters have been mostly determined by global
adjustments to various experimental data across the nuclear chart. It is extremely
gratifying to see that these adjustments of a few parameters give a fair
overall agreement with a large body of data. However, through these
adjustments, a precise description of particular experimental features
in a given region of nuclei, such as the transuranium elements, has not been obtained.

The most advanced adjustments of parameters with full error analyses have been
performed for the Skyrme energy-density functionals
(EDFs)~\cite{(Klu09),(Kor10b),(Kor12)}. It is by now quite clear that within a
limited parametrization of this force, further improvements towards globally
precise results are not possible~\cite{(Kor08),(Kor13)}. Much less work in this
direction has been done for the relativistic approaches and almost none for the
Gogny interaction and microscopic-macroscopic models, however, one can
reasonably expect that similar conclusions may hold there too.

Various strategies of extending simple parametrizations are now being
investigated. However, before these become available and eventually achieve
a higher level of precision, one has to look into the possibility of local
adjustment of parameters. This is the strategy adopted in the present work.
In doing so, the idea of Ref.~\cite{(Zal08)} is followed, where so-called
spectroscopic-quality parametrizations, which focus on adjusting the
shell-structure properties of nuclei, were proposed.

The paper is organized as follows. In Sec.\ref{sec2} the parameters
of the model are defined and various adjustments of parameters are discussed. In
Sec.\ref{sec4} the results obtained are presented and analyzed.
Finally, in Sec.\ref{sec5} the conclusions that can be drawn from the
study are presented.

\section{The model}
\label{sec2}

In this study, calculations were performed using the symmetry-unrestricted
solver {\sc hfodd} (v2.52j)~\cite{(Sch12),[Sch14]}. The Skyrme Hartree-Fock-Bogolyubov
(SHFB) equations were solved by expanding the mean-field wave functions on 680
deformed harmonic-oscillator (HO) basis states, with HO frequencies of
$\hbar \Omega_x = \hbar \Omega_y = 8.4826549$\,MeV and $\hbar \Omega_z =
6.4653456$\,MeV. This corresponds to including the HO basis states up to
$N_x=N_y=13$ and $N_z=14$ quanta. To study any possible impact of the
finite HO space, calculations were also performed using a larger
basis of 969 states and $N_x=N_y=15$ and $N_z=19$ quanta.

In odd-mass nuclei, quasiparticle excitations are obtained by blocking the relevant
levels when performing the SHFB calculations. The procedure closely follows
that of Ref.~\cite{(Sch10)}. The excitation spectra are obtained by taking
differences of the total energies obtained by blocking different orbitals.

The recently proposed Skyrme parametrization UNEDF1~\cite{(Kor12)} was used, for
which local adjustments of the spin-orbit (SO) strength were performed, while keeping all other
parameters unchanged (see below). The parameter set obtained is denoted
by UNEDF1$^{\text{SO}}$. For comparison, the results obtained using an older standard Skyrme parametrization
SLy4~\cite{(Cha98)} were calculated. The final calculations were performed for the force
UNEDF1$^{\text{SO}}_{\text{L}}$, for which the time-odd coupling constants were
determined by the Landau parameters and local-gauge-invariance arguments, as
defined in Ref.~\cite{(Ben02)}.

In the pairing channel, the mixed pairing force~\cite{(Dob02c)} was adopted,
both with and without Lipkin-Nogami (LN) approximate particle-number
projection~\cite{(Sto03)}. These two variants of the calculation are denoted by
SHFB+LN and SHFB, respectively. In both cases, the strengths of the pairing
force for neutrons ($V_0^{n}$) and protons ($V_0^{p}$) were adjusted either to
the odd-even mass staggering or first (kinematic) moments of inertia (MoI) $\cal{J}$$^{(1)}$.
Experimental and theoretical pairing gaps were estimated by using the
three-point-mass-difference formula, which takes into account the influence of
the deformation and blocking effects~\cite{(Sat98a),(Xu99w),(Dob01b)}.

\begin{figure}
\centering
\includegraphics[width=0.8\columnwidth]{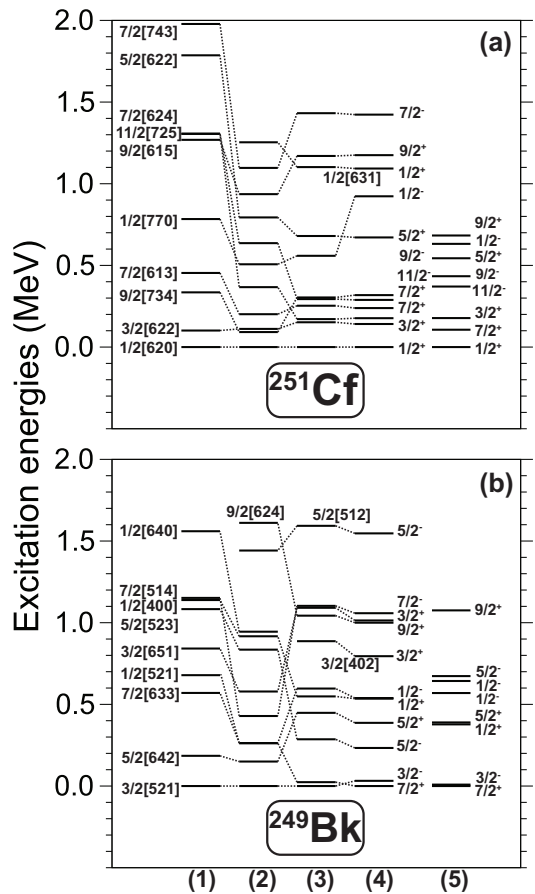}
\caption{The SHFB+LN excitation spectra of $^{251}$Cf (top) and $^{249}$Bk
(bottom). Results obtained with the Skyrme EDFs SLy4 (1), UNEDF1 (2),
UNEDF1$^{\text{SO}}$ (3), and UNEDF1$^{\text{SO}}_{\text{L}}$ (4) are
compared with experimental data (5) taken from the ENSDF
database~\cite{[ensdf]} and Refs~\cite{(Ahm05),(Ahm05b)}.
}
\label{fig3}
\end{figure}

Experimentally, the most detailed knowledge of the s.p.\ levels in the transuranium
region comes from the energy spectra of $^{251}$Cf and
$^{249}$Bk~\cite{(Her08)}. Therefore, the SHFB+LN
blocking calculations were performed for these two nuclei and the excitation spectra obtained
were compared with experimental data, see Fig.~\ref{fig3}. In general, it is found
that for SLy4 and UNEDF1, correct orbitals appear near the Fermi surface,
however, the detailed agreement is not fully satisfactory. It should be noted that
with the most recent parametrization of Ref.~\cite{(Was12)}, this situation
does not improve. The most conspicuous differences between the data and
calculations relate to positions of the intruder levels, which are highly
dependent on the SO coupling constants $C_0^{\nabla J}$ and $C_1^{\nabla J}$
of the Skyrme EDFs. The possibility of reproducing the experimental spectra in
odd-mass nuclei by changing the SO strengths was already discussed in
Ref.~\cite{(Ben03d)}.

In this work, focus was placed on the excitation energies of the $11/2^-$ state in
$^{251}$Cf with respect to the $1/2^+$ ground state and of the $3/2^-$ state in
$^{249}$Bk that is very close to the $7/2^+$ ground state. In theory, they
correspond to the relative energies between the neutron ($11/2^-[725]$) and proton
($7/2^+[633]$) intruder levels, and the corresponding normal-parity states
$1/2^+[620]$ and $3/2^-[521]$, respectively. The adjustment of parameters was
performed by repeating the following two steps:

\begin{itemize}
\item By assuming a linear dependence of the neutron and proton
excitation energies described above on the two coupling constants
$C_0^{\nabla J}$ and $C_1^{\nabla J}$, the experimental and theoretical
excitation energies of the $11/2^-$ state in $^{251}$Cf and $3/2^-$
state in $^{249}$Bk were matched.

\item By assuming a linear dependence of the neutron and proton mass staggering
on the two pairing strengths, $V_0^{n}$ and $V_0^{p}$, the
experimental and theoretical values obtained from the mass tables of Ref.~\cite{(Aud03)}, that
is, $\Delta^{(3)}_n=532.1$ and $\Delta^{(3)}_p=567.7$\,keV were matched.

\end{itemize}
In reality, it was found that the aforementioned dependencies were not strictly
linear, and that the two steps were not really independent from one
another, such that the procedure had to be repeated a few times, each time
iterating toward the final desired values. For the SHFB+LN calculations the final values read
\begin{eqnarray}
\label{SO}
(C_0^{\nabla J}, C_1^{\nabla J}) &=& (-88.050, 8.458)\,\mbox{MeV\,fm$^5$},\\
\label{VnVp}
(V_0^{n}, V_0^{p}) &=& (-191.1, -235.3)\,\mbox{MeV}.
\end{eqnarray}
The SO coupling constants in Eq.~(\ref{SO}) define the
parameterization UNEDF1$^{\text{SO}}$. Since four theoretical
parameters were adjusted to reproduce four experimental data points,
no estimates of the error in the adjusted values are possible. However,
the readjusted values (\ref{SO}) and  (\ref{VnVp}) are still within
2--3\,$\sigma$ deviations from the UNEDF1 values (see
the uncertainties $\sigma$ that were estimated in Ref.~\cite{(Kor12)}). In
fact, it may very well be that a full readjustment of the UNEDF1
parameter set, performed with constraints on energy levels in the
nobelium region, could result in a fit of a similar quality to that
obtained for the original force.

\begin{figure}
\centering
\includegraphics[width=0.9\columnwidth]{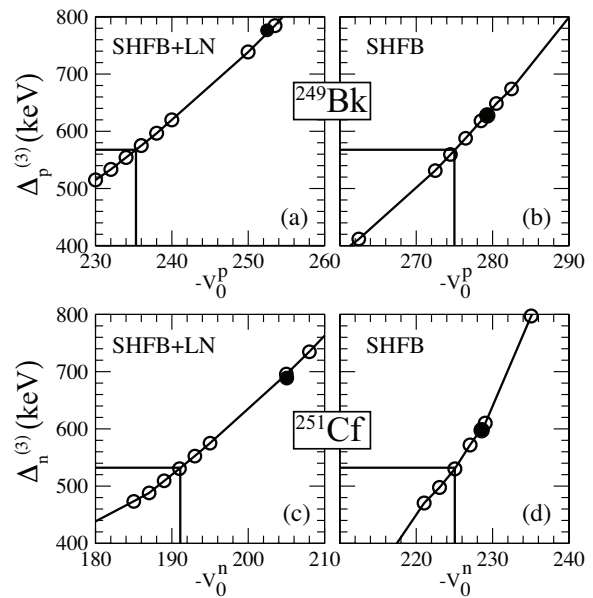}
\caption{The UNEDF1$^{\text{SO}}$ three-point mass differences $\Delta^{(3)}$
for protons in $^{249}$Bk
(top panels) and neutrons in $^{251}$Cf (bottom panels) as functions of
the proton and neutron pairing strengths, $V_0^p$ and  $V_0^n$, respectively.
Left and right panels show the SHFB+LN and SHFB results, respectively.
Vertical lines and full dots indicate values of the pairing strengths that give
experimental values of $\Delta^{(3)}$ and $\cal{J}$$^{(1)}$, respectively.
}
\label{fig1}
\end{figure}

Fig.~\ref{fig1} (left panels) shows the dependence of the calculated SHFB+LN
UNEDF1$^{\text{SO}}$ neutron and proton gaps on
the pairing strengths $V_0^n$ and $V_0^p$, respectively, with other
parameters fixed at the values shown in Eqs.~(\ref{SO}) and~(\ref{VnVp}).
Right panels show the analogous SHFB results.

\begin{figure}
\centering
\includegraphics[width=0.9\columnwidth]{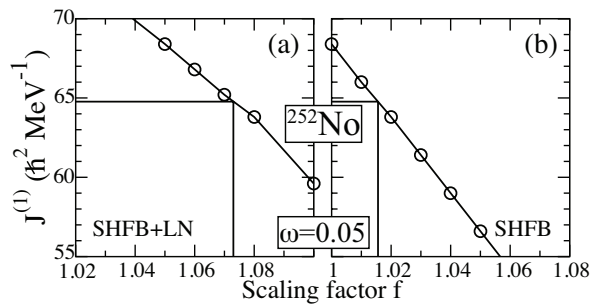}
\caption{Similar as in Fig.~\protect\ref{fig1}, but for the kinematic
MoI $\cal{J}$$^{(1)}$, calculated in $^{252}$No at the
rotational frequency of $\omega=0.05$\,MeV, as functions of the
pairing-strength scaling factors $f$, see text.
}
\label{fig2}
\end{figure}

With parametrization UNEDF1$^{\text{SO}}_{\text{L}}$ and pairing strengths
of Eq.~(\ref{VnVp}),  cranking SHFB+LN and SHFB
calculations for $^{246,248,250}$Fm, $^{252,254}$No, and $^{256}$Rf were performed.
In these nuclei, rotational bands have been measured experimentally. It was observed
that, at lower frequencies, the calculations overestimate
the measured kinematic MoI, $\cal{J}$$^{(1)}$, as shown in
Fig.~\ref{fig2} for $^{252}$No. The discrepancy is systematic and thus
calls for a generic explanation, which most probably should be
related to the overall strength of the pairing correlations. It means
that at the level of precision aimed at, the pairing strengths
inferred from the odd-even staggering and rotational properties are
incompatible.

Therefore, in order to better compare relative values of the MoI in
different nuclei, the pairing strengths $V_0^{n,p}$ of
Eq.~(\ref{VnVp}) were scaled by a common factor $f$, so as to match the
experimental value of $\cal{J}$$^{(1)}$ in $^{252}$No at $\omega=0.05$\,MeV
(see Fig.~\ref{fig2}).
Both for the SHFB+LN and SHFB calculations, the scaling gives values
slightly larger than one, namely $f=1.073$ and 1.017,
respectively, that is,
\begin{eqnarray}
\label{VnVpf1}
(V_0^{n}, V_0^{p}) &=& (-205.1, -252.5)\,\mbox{MeV,~ for SHFB+LN}, \\
\label{VnVpf2}
(V_0^{n}, V_0^{p}) &=& (-228.6, -279.3)\,\mbox{MeV,~ for SHFB}.
\end{eqnarray}
This, in turn, leads to an increase of the calculated pairing gaps
$\Delta^{(3)}_{n,p}$ by 150-200\,keV, as shown by dots in
figure~\ref{fig1}. At this point, it is very difficult to say if the
differences in pairing strengths, needed to describe the odd-even mass
staggering in $^{251}$Cf and $^{249}$Bk and the MoI in $^{252}$No is
significant. Both the gaps and the MoI are not only sensitive functions of
pairing, but also depend on other fine details of the shell
structure, which still can be imperfect, even after our careful
adjustments of the SO strengths. Finally, for consistency, all
results presented in this study were obtained with scaled pairing
strengths of Eqs.~(\ref{VnVpf1}) and (\ref{VnVpf2}). This includes
all SHFB+LN excitation spectra shown in Fig.~\ref{fig3}.

Tests performed for a larger HO basis of 969 states allow
the precision of the results obtained to be estimated. Firstly, it is noted that
the pairing properties do depend on the size of the HO space. For
example, adjustments of the pairing strengths to odd-even mass
staggering lead in the larger basis to values of $V_0^{n}$ smaller by
about 1\% (SHFB+LN) or 5\% (SHFB). At the same time, values of
$V_0^{p}$ are almost unaffected. This shows that values of the
pairing strengths given in Eqs.~(\ref{VnVp})--(\ref{VnVpf2}) pertain
to the specific size of the HO basis and cannot be considered as having
any universal significance. However, when all adjustments of
parameters are performed in the basis of 969 states, exactly in the
same way they were performed for the 680 states, final values of MoI
stay the same within 1\,$\hbar^2/$MeV.

\begin{figure}
\centering
\includegraphics[width=0.9\columnwidth]{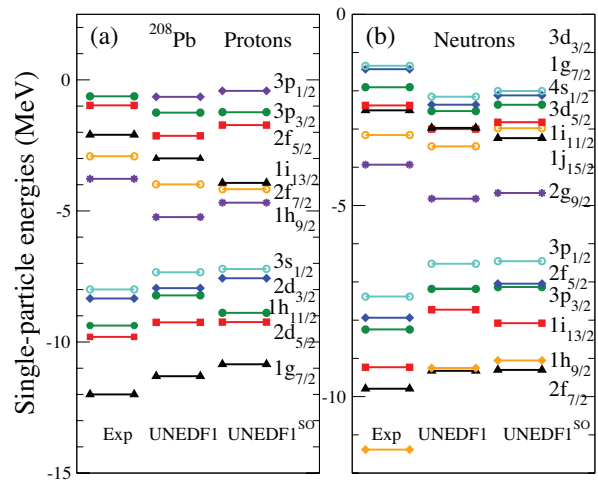}
\caption{(Color online) Proton (left) and neutron (right) s.p.\ spectra in $^{208}$Pb,
calculated for the UNEDF1 and UNEDF1$^{\text{SO}}$ functionals and
compared with evaluated experimental data~\cite{(Sch07)}.
}
\label{lead208}
\end{figure}

\begin{figure}
\centering
\includegraphics[width=0.9\columnwidth]{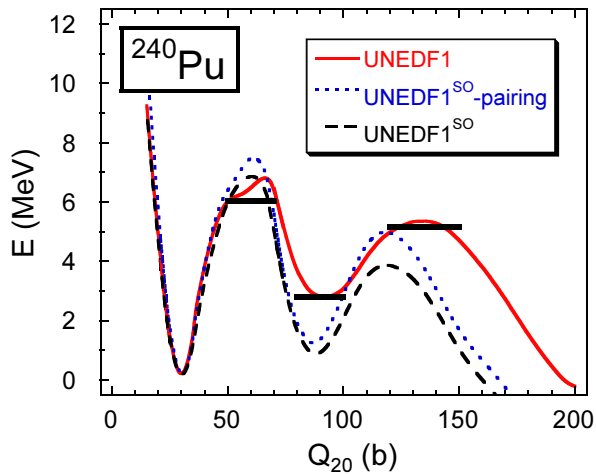}
\caption{(Color online) Fission barriers of $^{240}$Pu, calculated for the UNEDF1 (solid line)
and UNEDF1$^{\text{SO}}$ (dashed line) functionals. The dotted line shows
results for the hybrid functional using the UNEDF1$^{\text{SO}}$ SO strengths
and the original UNEDF1 pairing strengths. Horizontal lines illustrate the
experimental energies as explained in Ref.~\cite{(Kor12)}.
}
\label{pu240}
\end{figure}

To further illustrate the degree of changes induced by modifying the
SO and pairing parameters of the UNEDF1 functional, Figs.~\ref{lead208} and~\ref{pu240}
show results obtained for the s.p.\ spectra in $^{208}$Pb and fission barriers of $^{240}$Pu,
respectively. Here, identical numerical conditions and
experimental data as those used in Ref.~\cite{(Kor12)} are employed. It is seen
that the $^{208}$Pb s.p.\ energies obtained within the UNEDF1 and
UNEDF1$^{\text{SO}}$ functionals are very similar to one another,
with slightly larger SO splittings of UNEDF1$^{\text{SO}}$ being
in somewhat better agreement with data. However, it should be noted that the
overall discrepancies with respect to experimental data are significantly larger than
the differences between them for both functionals.

For the $^{240}$Pu fission barriers, the results depend significantly  on
the strength of the pairing correlations. Therefore, in
Fig.~\ref{pu240} an additional dotted line shows the results obtained for the
modified SO strengths of the UNEDF1$^{\text{SO}}$ functional, but
with the original UNEDF1 pairing strengths. It can be seen that at the top of barriers,
the differences in pairing alone create significantly different energies. However,
independently of the pairing strengths, the UNEDF1$^{\text{SO}}$
excitation energies of the second minima turn out to be significantly
underestimated. This shows that in a global fit of the
UNEDF1$^{\text{SO}}$ parameters, further readjustment of surface
energy may still be required. At this point, it cannot be expected that
after re-adjusting the spectroscopic SO properties, the detailed bulk nuclear
energies should stay the same.

\section{Results}
\label{sec4}

With the readjusted SO and pairing parameters, Eqs.~(\ref{SO})
and~(\ref{VnVpf1}), the UNEDF1$^{\text{SO}}$ and UNEDF1$^{\text{SO}}_{\text{L}}$
excitation spectra of $^{251}$Cf and $^{249}$Bk agree quite well with the experimental data, see
Fig.~\ref{fig3}. The set UNEDF1$^{\text{SO}}_{\text{L}}$, based on the Landau
parameters, gives energy levels that are only slightly different, by
100--200\,keV, than those of the UNEDF1$^{\text{SO}}$ set. The only exception
is the $1/2^-[770]$ state in $^{251}$Cf, which differs by about 400\,keV.

For the neutron spectrum, the low-energy part is perfectly reproduced. The only
more significant discrepancy between calculations and experiment is that there
are two $7/2^+$ states obtained close to one another, whereas only one is seen
in the data. The Nilsson numbers of these two states are $7/2^+[613]$ and
$7/2^+[624]$, and they have particle and hole character, respectively. At
higher energies, three levels, $5/2^+[622[$, $1/2^-[770]$, and $9/2^+[615]$, are more
stretched compared to the experimental data, but their order is correctly
reproduced.

Throughout the paper we use the standard concept of the Nilsson
asymptotic quantum numbers~\cite{(Nil55),(RS80)}, assigned to
deformed s.p.\ states. Although these numbers are exact only in the
limit of large deformations of the axial HO, based on determining the
largest Nilsson components of deformed wave functions they are
customarily assigned to all s.p.\ states. This creates a robust
platform of describing various deformed configurations, which is most
often valid across different models, and which is also routinely used
in the discussions of experimental properties of deformed nuclei.
This may create ambiguities only when more than one large Nilsson
component is present in a given model wave function. In
this context, we note that the Nilsson quantum numbers for $1/2^-$
state in our SHFB+LN calculation are $1/2^-[770]$, whereas in the
analyses of experimental data~\cite{(Ahm05b)}, they are given as
$1/2^-[750]$.

\begin{figure}
\centering
\includegraphics[width=0.8\columnwidth]{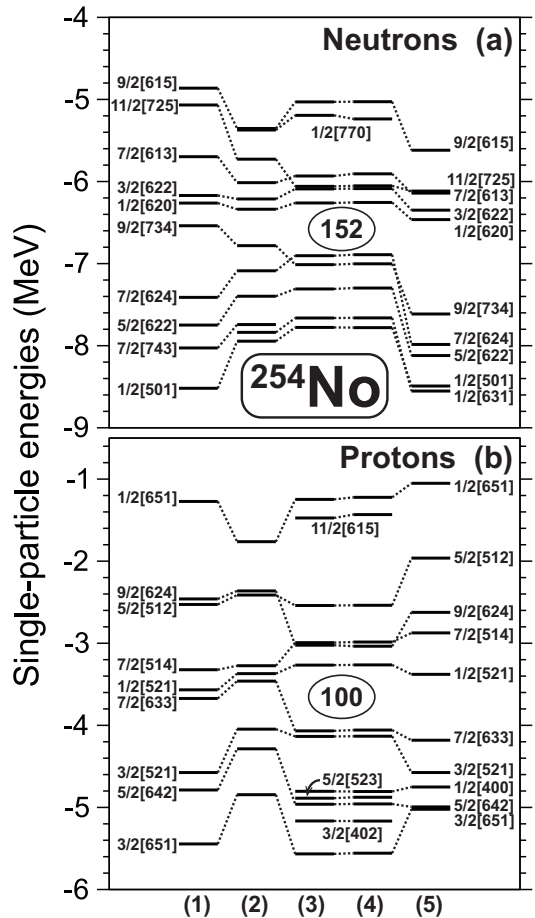}
\caption{Single-particle energies of neutrons (top) and protons (bottom) in
$^{254}$No. Columns (1)--(4) are labeled as in Fig.~\protect\ref{fig3},
whereas column (5) gives results calculated for the Woods-Saxon
potential~\protect\cite{(Cwi87)}.
}
\label{fig4}
\end{figure}

It is especially gratifying to see the proton $7/2^+[633]$ level almost degenerate
with the $3/2^-[521]$. On the one hand, within the standard SO strengths, such a
degeneracy is very difficult to obtain. On the other hand, in experiment it is
a recurring feature of odd-proton nuclei in this region~\cite{(Hes05),(Shi14)}.
The rest of the low-energy spectrum is also in good agreement with experiment,
with only the $5/2^-[523]$ state appearing lower in calculation than in experiment.
This suggests that the proton 1h$_{9/2^-}$ spherical shell may have been
brought too close to the Fermi level. An additional $1/2^-[530]$ state seen
in experiment is not present in the calculations~\cite{(Ahm05)}.

The good agreement between the calculated and experimental spectra of odd-mass nuclei,
prompted an analysis of the s.p.\ deformed shell structure resulting from the
calculations. Although the deformed s.p.\ levels are not directly observable,
they still nicely illustrate basic features of paired deformed systems. In
Fig.~\ref{fig4}, the s.p.\ levels of $^{254}$No obtained for the
EDF parameterizations that were shown in Fig.~\ref{fig3} are compared with those calculated
for the Woods-Saxon potential~\cite{(Cwi87)} with deformation parameters taken
from Ref.~\cite{(Sob01)}.

\begin{figure}
\centering
\includegraphics[width=\columnwidth]{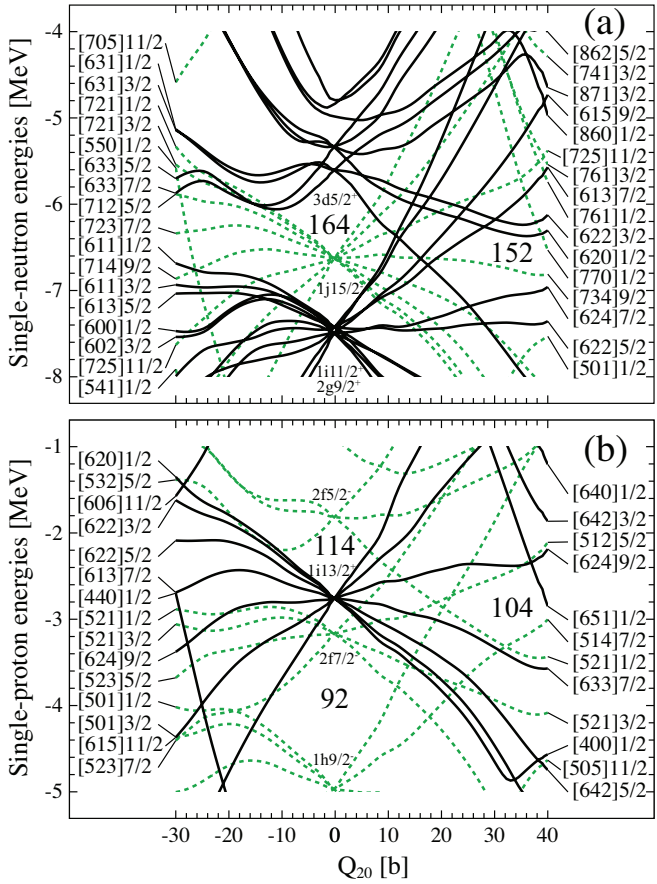}
\caption{(Color online) Neutron (top) and proton (bottom) Nilsson diagrams in
$^{254}$No calculated for the Skyrme EDF UNEDF1. Even- and odd-parity levels
are drawn with solid and dashed lines, respectively.
}
\label{fig5}
\end{figure}

\begin{figure}
\centering
\includegraphics[width=\columnwidth]{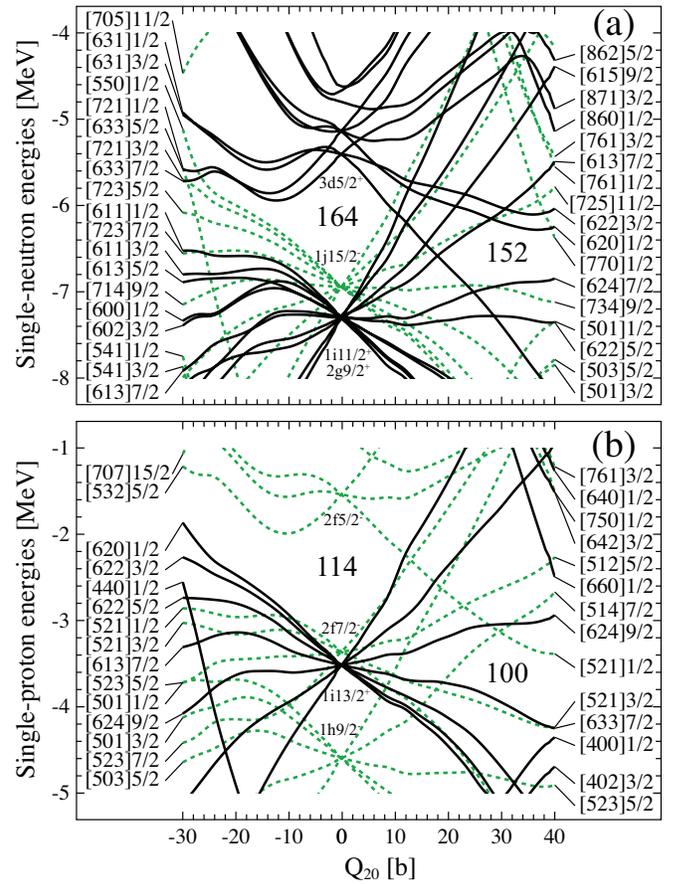}
\caption{(Color online) Same as in Fig.~\protect\ref{fig5} but for the Skyrme EDF
UNEDF1$^{\text{SO}}$ derived in this work.
}
\label{fig6}
\end{figure}

For the SLy4 EDF, the neutron and proton shell gaps open up at $N=150$ and
$Z=98$ and $104$, cf.~Ref.~\cite{(Ben13)}. The original UNEDF1 parameterization
gives more compressed s.p.\ spectra, with no apparent neutron shell gap and
the proton shell gaps also appearing at $Z=98$ and $104$. With the adjusted SO
terms, the s.p.\ energies calculated with UNEDF1$^{\text{SO}}$ show the shell
gaps that open up at $N=152$ and $Z=100$. This result is very similar to the
one obtained for the phenomenological Woods-Saxon potential, which in this
region is considered to be consistent with experiment.

Figures~\ref{fig5} and \ref{fig6} display for the UNEDF1 and
UNEDF1$^{\text{SO}}$ parametrizations, respectively, the s.p.\ energies as
functions of the quadrupole moment (Nilsson diagrams). It is seen that the
readjustment of the SO coupling constants results in the intruder neutron
1j$_{15/2^-}$ and proton 1i$_{13/2^+}$ spherical shells being shifted down by a
few hundred keV with respect to the normal-parity levels. This, at deformation
of $Q_{20}\approx 33$\,b, opens up the deformed neutron $N=152$ and proton
$Z=100$ gaps. At the same time it can be seen that the UNEDF1$^{\text{SO}}$ parameters
give the spherical proton-shell opening at $Z=114$. It is noted that the issue
of spherical shell gaps of superheavy nuclei, highly debated in the literature, in fact
depends on a quite tiny readjustment of two poorly determined
parameters of the underlying theory.

\begin{figure*}
\centering
\includegraphics[width=0.7\textwidth]{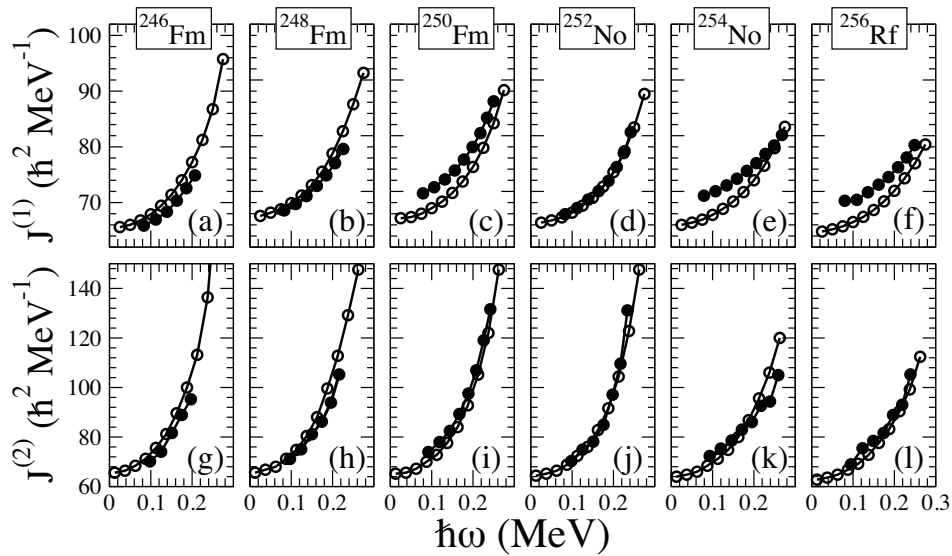}
\caption{The kinematic $\cal{J}$$^{(1)}$ (top) and dynamic $\cal{J}$$^{(2)}$ (bottom)
MoI in $^{246}$Fm, $^{248}$Fm, $^{250}$Fm,
$^{252}$No, $^{254}$No, and $^{256}$Rf as functions of the rotational
frequency. Open circles show theoretical results whereas full dots
denote experimental values.
}
\label{fig7}
\end{figure*}

In Fig.~\ref{fig7}, the calculated (cranked SHFB+LN with
UNEDF1$^{\text{SO}}_{\text{L}}$) and experimental, kinematic ($\cal{J}$$^{(1)}$)
and dynamic ($\cal{J}$$^{(2)}$) MoI are shown for six nuclei where experimental values
are available. As discussed in Sec.~\ref{sec2}, increased pairing strengths (\ref{VnVpf1}), adjusted to the
experimental value of the $\cal{J}$$^{(1)}$ MoI at $\omega=0.05$ in
$^{252}$No are used. In the scale of Fig.~\ref{fig7}, the
agreement between the calculations and experiment is reasonably good.
Especially the dynamic MoI $\cal{J}$$^{(2)}$ are reproduced almost perfectly.
However, in this work focus is placed on the fine details of the
rotational alignment, which may give hints concerning
the underlying shell structure.

\begin{figure}
\centering
\includegraphics[width=\columnwidth]{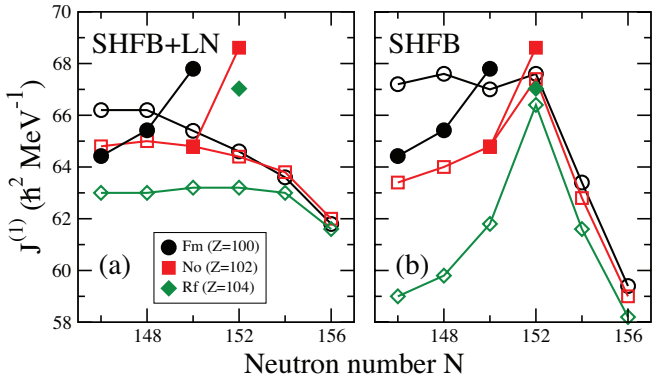}
\caption{(Color online) The kinematic MoI $\cal{J}$$^{(1)}$
of the Fm, No, and Rf isotopes calculated at the rotational frequency
of $\omega=0.05$\,MeV. Left and right panels show the SHFB+LN and
SHFB results. Open and full symbols show theoretical results and
available experimental data, respectively.
}
\label{fig8}
\end{figure}

Therefore, in Fig.~\ref{fig8} an extended scale is used to compare
the $\omega=0.05$\,MeV calculated and experimental values of
$\cal{J}$$^{(1)}$. The latter were obtained by the standard Harris fits to
experimental data~\cite{(Gre12)}. Cranking calculations were
performed for 18 even-even nuclei with $Z=100$--104 and $N=146$--156,
with (SHFB+LN) and without (SHFB) the LN corrections. The
SHFB+LN results show a rather smooth decrease of $\cal{J}$$^{(1)}$ with
increasing neutron number, whereas those obtained for SHFB show at
neutron number $N=152$ pronounced peaks for No and Rf isotopes, or a
kink for the Fm isotopes, giving a much better agreement with data.
It is noted that similar trends were obtained with and without LN
corrections, respectively, in Refs.~\cite{(Afa13)} and~\cite{(Sob01)}.
These studies used different, relativistic and phenomenological,
p-h mean-fields, but did not compare results obtained with and
without LN corrections.

The appearance of a peak of $\cal{J}$$^{(1)}$ may indicate that the Fermi level
reaches the s.p.\ shell gap. This results in a quenching of pairing
correlations, and thus in an increase of the MoI. The LN correlation,
 as in any other method to restore the particle-number
symmetry, tends to enhance pairing effects, even at shell gaps.
Therefore, such methods lead to smoother dependencies of pairing on
particle numbers and thus wash out sudden changes in the MoI. It seems that
this mechanism leads, at variance with experiment, to a complete
disappearance of the effects of the $N=152$ shell gap on rotational
properties.

\section{Conclusion}
\label{sec5}

In the present work, the rotational properties
of very heavy nuclei have been investigated through the magnifying glass of a
spectroscopic-quality energy density functional, which was
specifically adjusted to this region of nuclei. This approach is
complementary to numerous studies performed using globally
adjusted functionals, where gross features are usually nicely
reproduced, but finer details, which are of great interest to
experimentalists, are often not.

Indeed, by adjusting the SO coupling constants and pairing strengths
to odd-even mass staggering and intruder excitation energies in
$^{251}$Cf and $^{249}$Bk, it has been possible to obtain good
agreement with data. Moreover, the obtained deformed shell structure
then revealed pronounced shell gaps, which opened up at $Z=100$ and $N=152$,
again in agreement with general features of the experimental data
in this region of nuclei. Cranking calculations performed with the
modified parameters also gave a fair agreement with experiment.
However, under detailed comparison at a much finer level than usually performed,
the following deficiencies of the description were revealed:

\begin{itemize}
\item
The pairing strengths needed for the precise description of the
odd-even mass staggering and those needed for a precise description
of the moments of inertia are not exactly the same.

\item
The approximate particle-number projection, performed using the
Lipkin-Nogami method, gives a dependence of moments
of inertia on particle numbers which is too smooth. It seems that the particle-number
restoration washes out the deformed shell gaps, which is not compatible
with data. This is puzzling, as in principle the particle-number conserving
approach should give a superior description of the
experimental data.

\end{itemize}

It is suggested that, within the current restricted parametrizations of the
energy density functionals, further global improvement of precision
in reproducing experimental data is unlikely. Of course, much more
work is needed before extended parametrizations can be implemented and
tested. In the mean time, local focused adjustments, like the one
performed in this study, can give hints as to which elements of the
description need more attention than others. It is hoped that the
two deficiencies identified in this work may help in the future
quest for precision.

\begin{acknowledgments}
This work was supported in part by the Academy of Finland and
University of Jyv\"askyl\"a within the FIDIPRO programme, Centre
of Excellence Programme 2012--2017 (Nuclear and Accelerator Based
Physics Programme at JYFL), European Research Council through the
SHESTRUCT project (grant agreement number 203481), Office
of Nuclear Physics,
U.S. Department of
Energy (DOE) under Contracts No.
DE-FG02-96ER40963
(University of Tennessee),  No.
DE-SC0008499    (NUCLEI SciDAC Collaboration),
and Polish National Science Center under Contract
No.\ 2012/07/B/ST2/03907. The CSC - IT Center for
Science Ltd, Finland, is acknowledged for the allocation of computational resources.
\end{acknowledgments}

\bibliographystyle{apsrev4-1}
%

\end{document}